\documentstyle[prb,twocolumn,aps,psfig]{revtex}

\draft
\begin{document}
\twocolumn[\hsize\textwidth\columnwidth\hsize\csname@twocolumnfalse%
\endcsname

\title{Phase Diagram of the ${\bf {S=}} {\textstyle \frac{\bf 1}{\bf 2}}$ 
Frustrated Coupled Ladder System}

\author{B. Normand$^1$, K. Penc$^2$, M. Albrecht$^3$ and F. Mila$^3$}

\address{$^1$Theoretische Physik, ETH-H\"onggerberg, CH-8093 Z\"urich, 
Switzerland. }

\address{$^2$Max Planck Institut f\"ur Physik komplexer Systeme, 
Bayreuther Str. 40, D-01187 Dresden, Germany. }

\address{$^3$Laboratoire de Physique 
Quantique, Universit\'e Paul Sabatier, 118 Route de Narbonne, 31062 
Toulouse Cedex, France. }

\date{\today}

\maketitle

\begin{abstract}

	We present a theoretical study of the magnetic phase diagram of the
frustrated coupled ladder structure realized recently in several materials. 
This system displays a nondegenerate spin-gap state in the dimer limit and 
an infinitely degenerate spin-gap state in the regime of weakly-coupled 
zig-zag chains. Between these we demonstrate the existence of gapless, 
magnetically ordered regions whose order is antiferromagnetic close to the 
honeycomb lattice limit, and incommensurate along the chains when all three 
magnetic interactions compete. 

\end{abstract}

]

%\section{Introduction}

        Spurred by developments in the field of high-temperature 
superconductivity, rapid progress is now being made in the preparation of 
materials with similar attributes, and the potential for fascinating new 
physics. These are inorganic, low-dimensional quantum magnets, where the 
relative strengths of the magnetic interactions, which in most cases give 
antiferromagnetic coupling between ions with spin $S = {\textstyle 
\frac{1}{2}}$, result in systems such as chains and ladders, which are 
effectively one-dimensional (1d), or 2d coupled ladders and depleted planes. 
One structure of particular interest is the frustrated coupled ladder, shown 
in Fig. 1. This conformation is realized in the cases of the prototypical 
``ladder'' compound \cite{rhatb} SrCu$_2$O$_3$, where the frustrating 
interaction $J_1$ is ferromagnetic (FM) and small, so that the 
$J_2$-$J_2^{\prime}$ ladders are only weakly coupled, and in 
the ``zig-zag'' (or $J_1$-$J_2$) chain compound \cite{rhatb} SrCuO$_2$, where 
the interchain coupling $J_2^{\prime}$ is small. The depleted planar compound 
\cite{rbg} CaV$_2$O$_5$ represents a system where all three interactions are 
thought to be of similar strength. 

\begin{figure}[hp]
\centerline{\psfig{figure=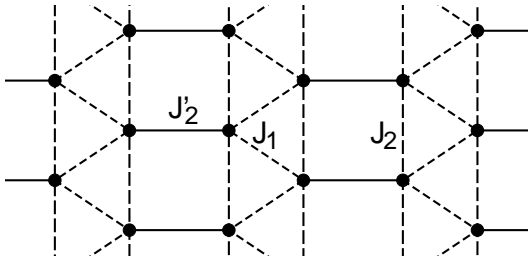,height=3.5cm,angle=0}}
\medskip
\caption{Schematic representation of the frustrated coupled ladder system. }
\end{figure}

	The phase diagram obtained on alteration of the parameters $J_1$, 
$J_2$ and $J_{2}^{\prime}$ is particularly rich. The two-chain ladder obtained
at $J_1 = 0$ has a singlet ground state of resonating dimers with a gap to all
spin excitations, while the zig-zag chain at $J_2^{\prime} = 0$ shows for 
$J_2 > J_{2c}$ a doubly degenerate ground state of alternating dimers which 
also has a gap. What appears between these limits remains poorly understood, 
and is one subject of the current analysis, but we may gain an initial 
indication from the honeycomb lattice obtained when $J_2 = 0$, where the 
gapped $J_2^{\prime}$ dimer phase is replaced by a magnetically ordered state 
at a critical value of increasing $J_1$. 
We investigate the system using a variety of analytical and numerical methods,
proceeding in (a) from the limit of ladder rung dimers, in (b) from that of
weakly coupled zig-zag chains and in (c) by considering the ordered phases 
which are found to occur between these limits. 

%\section{Dimer Limit}
\paragraph{Dimer Limit:}
	in the limit of large $J_{2}^{\prime}$, the system has a 
non-degenerate ground state, whose wavefunction is a product of rung dimer
singlets, with an energy gap to spin excitations. This state is well suited to 
examination by the bond-operator technique, \cite{rsb} which is based on 
transforming the four spin states on each rung to one singlet and three 
triplets. The gap between these gives the stability of dimer order. The 
method has also been shown \cite{rgrs} to be applicable to spin ladders
(vanishing $J_1$) for values of $J_2$ up to $J_{2}^{\prime}$, and can 
be expected to have similar validity in the case of the anisotropic honeycomb 
lattice (vanishing $J_2$). 

	Following the treatment of Ref. \onlinecite{rnr} for
a unit cell containing two dimers, and retaining in the spin-gap regime terms 
only to quadratic order in the triplet spin excitations $t_{i \alpha}^{\dag}$,
we obtain a system of threefold degenerate magnons with energy spectrum 
\begin{equation}
\omega_{\bf k}^{\pm} = \left( {\textstyle \frac{1}{4}} J_2^{\prime} 
- \mu \right) \sqrt{ 1 + d a_k^{\pm} } ,
\label{ems}
\end{equation}
where $d = 2 J_2^{\prime} {\overline s}^2 / \left( {\textstyle \frac{1}{4}} 
J_2^{\prime} - \mu \right)$,
\begin{equation}
a_{\bf k}^{\pm} = \lambda \cos k_z \pm \lambda^{\prime} \cos 
{\textstyle \frac{1}{2}} k_x \cos {\textstyle \frac{1}{2}} k_z , 
\end{equation}
$\overline s$ denotes the magnitude of the singlet condensate, $\mu$ is the
global chemical potential, and we define $\lambda = J_2 / 
J_2^{\prime}$ and $\lambda^{\prime} = J_1 / J_2^{\prime}$. Solution of the 
mean-field equations at $T = 0$ gives the most important property 
characterizing the non-degenerate ground state, the spin gap 
\begin{equation}
\Delta = \left( {\textstyle \frac{1}{4}} J_2^{\prime} - \mu \right) \sqrt{1 
- d \, f \left( \lambda, \lambda^{\prime} \right)} .
\label{esg}
\end{equation}

	For $J_1 > 4J_2$, $f = \lambda^{\prime} - \lambda = a_{k,{\rm min}}^-$
at the commensurate wave vector $k_{\rm M} = (0,0)$, while for $J_2 > 
{\textstyle \frac{1}{4}} J_1$ one has $f = \lambda + \frac{\lambda^{\prime 
2}}{8 \lambda}$ occurring at an incommensurate $k_{\rm M} = (0, 2 \cos^{-1} 
\frac{\lambda^{\prime}}{4 \lambda})$. In the spin-gap regime, the maximum of 
the static structure factor $S(k)$ coincides with the minimum of the gap, and 
can be shown to move to incommensurate wave vectors with increasing $J_2$ 
beyond ${\textstyle \frac{1}{4}} J_1$. The boundaries of the spin-gap phase 
are found where the $\Delta \rightarrow 0$, and are shown in Fig. 2. When 
$J_1 > 4J_2$, from $k_{\rm M}$ one may deduce a FM arrangement of rung 
spin pairs ($|\uparrow \downarrow \rangle$), which corresponds to 
antiferromagnetic (AF) order along the $J_1$ chains. For $J_2 > 
{\textstyle \frac{1}{4}} J_1$, $k_{\rm M}$ describes spiral order along 
the chains with exactly the pitch known from the classical solution for 
the $J_1$-$J_2$ chain. The phase boundaries in Fig. 2 appear well outside 
the regime of validity of the bond-operator technique, which overestimates 
the stability of the dimer state. However, they remain qualitatively correct,
particularly in showing the instability of the dimer liquid to the different 
ordered phases, and furthermore provide a useful indication of where 
to apply alternative approaches. 

\begin{figure}[hp]
\centerline{\psfig{figure=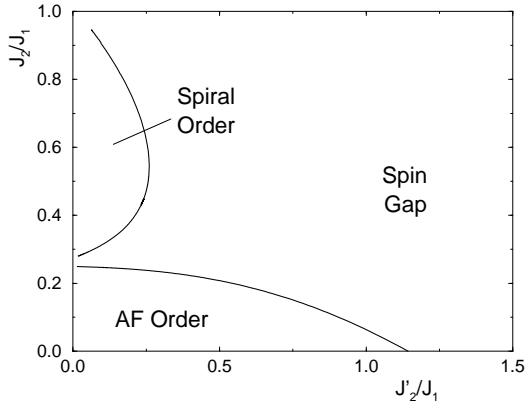,height=6.2cm,angle=270}}
\medskip
\caption{Phase diagram of the coupled dimer system, shown in the plane of 
$J_2^{\prime} / J_1$ and $J_2 / J_1$, as deduced from the bond-operator 
method. }
\end{figure}

%\section{Frustrated Chain Limit}
\paragraph{Frustrated Chain Limit:}
for large $J_1$ or $J_2$, the system consists of weakly coupled chains, 
whose properties are well known from the Bethe Ansatz solution. However, how 
these are affected by the weak couplings, which may or may not be frustrating,
is an extremely subtle issue to which we shall return below. Here we consider 
the limit of small $J_{2}^{\prime}$, where the problem is that of the 
$J_1$-$J_2$ chain.  This is a classic example of frustration in 1d, 
and we review briefly its properties (see Ref. \onlinecite{raw}, and
Refs. (12)-(18) therein). At the Majumdar-Ghosh (MG) point $J_2 = {\textstyle
\frac{1}{2}} J_1$, there is an exact dimer wave function solution 
\cite{rmg} with a twofold degenerate ground state and gapped excitation 
spectrum. \cite{rss} The wave function is composed of frozen singlets on 
alternating $J_1$ bonds, and has two possible realizations differing by a 
translation of one unit along the chain. The gap decreases exponentially 
with $J_2 - J_{2c}$ \cite{rh} on approach to the conformal point $J_{2c}/J_1 
= 0.2412$, below which the frustrated chain retains the gapless spectrum of 
the nearest-neighbor spin chain. For $J_2 > {\textstyle
\frac{1}{2}} J_1$ the gap increases to a maximum 
at $0.6 < J_2/J_1 < 0.7$, then tends towards an exponential decay  
with $-J_2/J_1$ for large $J_2$. The static spin-spin correlation function 
$S(k)$ \cite{rth} has a maximum at $k = \pi$ for $J_2 \leq {\textstyle 
\frac{1}{2}} J_1$, and for $J_2 > {\textstyle \frac{1}{2}} J_1$ is maximal 
between $k = {\textstyle \frac{1}{2}} \pi$ and $\pi$, indicating predominant 
spiral correlations. 

	We examine the lifting of degeneracy and the closing of the 
singlet-triplet gap $\Delta$ by an adapted perturbational approach, which due 
to the degenerate nature of the ground state is in a strict sense variational. 
The Hamiltonian is written as $H = H_0 + H^{\prime}$, where 
\begin{eqnarray}
H_0 & = & J_1 \sum_i {\bf \hat S}_{i,l} {\bf \hat S}_{i+1,l}
                    +J_2 \sum_i {\bf \hat S}_{i,l}{\bf \hat S}_{i+2,l} 
\label{eho} \\ H^{\prime} & = & J'_2 \sum_{i,l \rm ~even} 
                  {\bf \hat S}_{i,l} {\bf \hat S}_{i,l+1} + 
          J'_2 \sum_{i,l \rm ~odd} {\bf \hat S}_{i,l} {\bf \hat S}_{i,l+1} , 
\label{ehi}
\end{eqnarray}
and in $H^{\prime}$ both $i$ and $l$ take either even or odd values only. 
We construct the variational wave function
\begin{equation}
 |\Psi_j \rangle = |S\rangle_0 \otimes \dots \otimes 
|S\rangle_{j-1} \otimes |T\rangle_{j} \otimes|S\rangle_{j+1} \dots ,
\end{equation}
in which $|S\rangle_i = \eta |S_0\rangle + \eta' |S_\pi\rangle $ is a linear 
combination of the two degenerate ground states on the $i$-th chain.  
$|T\rangle_j = \sum_{\alpha} (\xi_{k,\alpha} |T_{k,\alpha}\rangle + 
\xi_{k+\pi,\alpha} |T_{k+\pi,\alpha}\rangle)$, where $|T_{k,\alpha}\rangle$ 
is the eigenstate with momentum $k$ of the isolated zig-zag chain, denotes 
a triplet excitation of the $j$-th chain. In the thermodynamic limit these 
form a continuum above the gap, and their delocalization by $H^{\prime}$ 
gives a kinetic energy gain which reduces $\Delta$. 

	In the regime $J_2 < {\textstyle \frac{1}{2}} J_1$ where the spin
correlations are predominantly antiferromagnetic, the lowest-lying triplet 
excitations appear at $k = 0$ and $k = \pi$. Here $\eta$ or $\eta' = 0$, 
and the ``gap equation'' for $J_2^{\prime}$ takes the simplified form 
\begin{equation}
  \frac{1}{J'_2} = 
    \max \sum_\alpha \left\{ 
       \frac{|\langle T_{\pi,\alpha}|\hat S^z_\pi|S_0 \rangle|^2}
            {\epsilon_\alpha-\epsilon_S-\Delta} ,
       \frac{|\langle T_{0,\alpha}|\hat S^z_\pi|S_\pi \rangle|^2}
            {\epsilon_\alpha-\epsilon_S-\Delta}
       \right\} ,
\end{equation}
where $\epsilon_S$ and $\epsilon_\alpha$ are the ground and excited state
energies, and may be evaluated in terms of the spin-spin correlation function 
\begin{equation}
{\rm Im } S^{zz}_{q}(\omega,k) = \sum_\alpha 
    |\langle T_{k+q,\alpha}|\hat S^z_k |S_q \rangle|^2
            \delta(\omega-\omega_{k+q}) ,
\end{equation} 
in which $\omega_{k+q} = \epsilon_{k+q,\alpha} - \epsilon_S$. These 
expressions are calculated using a L\'anczos algorithm for chain lengths 
up to $N = 22$ sites. Setting $\Delta = 0$ yields the value $J'_{2c}$ 
where the ordered phases appear. $J'_{2c}$ scales with the unperturbed gap
magnitude $\Delta_0$, and vanishes at $J_2 = J_{2c}$. For $J_2 > 
{\textstyle \frac{1}{2}} J_1$, the situation is more complicated because the 
lowest triplet excitations appear at intermediate momenta, corresponding 
to predominant incommensurate spin correlations. \cite{rth} There is no 
decoupling as above, necessitating solution of the full $J_2^{\prime}$ 
equation and calculation of the energy by minimizing with respect to $\eta$ 
and $\eta'$. $\Delta \rightarrow 0$ for a value of $J_{2}^{\prime}$ whose 
dependence on $\Delta_0$ varies because of the incommensuration, as shown in
the inset of Fig. 3. That gap closure occurs 
at a wave vector away from $\pi$ we take as an indication
of spiral magnetic order. The method of solution is most accurate 
where the gap of the dimerized phase is large, and we display in Fig. 3
the phase diagram for the parameter regime $0.4 < J_2/J_1 \leq 1$. 
The nature of the magnetic order stabilized by increasing $J_{2}^{\prime}$ 
corresponds to the wave vector $k_{\rm M}$ maximizing $S(k)$. \cite{rth}

\begin{figure}[hp]
\centerline{\psfig{figure=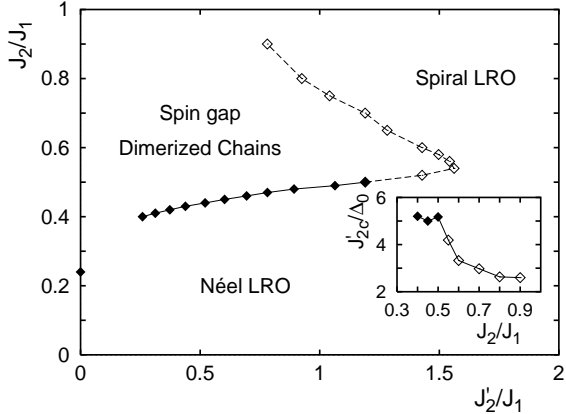,height=5.5cm,angle=0}}
\medskip
\caption{Phase diagram of the coupled zig-zag chain system, as deduced from
the variational method. Solid symbols denote the commensurate solution 
($J_2 < {\textstyle \frac{1}{2}} J_1$) and empty symbols the incommensurate.}
\end{figure}

	We may analyze the energy shift and splitting of the 
ground state by applying second-order degenerate perturbation theory. At this
level one need only consider two coupled chains, which have a four-fold
degenerate ground state, and the energy correction per site is a $4\times 4$ 
matrix conveniently calculated using the L\'anczos algorithm. The correction 
is found to be proportional to the unit matrix, so the degeneracy is not 
lifted in the thermodynamic limit to second order. This result suggests that 
the degeneracy of the ground state will be reflected in the presence of 
low-lying singlet states in the singlet-triplet gap for small $J_{2}^{\prime}$.
We note that this behavior contrasts with the lifting of degeneracy which
occurs in a system of gapped spin chains where $i$ and $l$ may take all 
values in Eq. (\ref{ehi}). 

%\section{Ordered Phases}
\paragraph{Ordered Phases:} 
these may be treated by expanding in fluctuations around
a state with fixed moments, of periodically varying orientation, on each 
site. We present here results from the Schwinger boson and linear spin 
wave methods. In the Schwinger boson transformation, \cite{raa,rmpb} ordered 
solutions are described by a Bose condensation where $\langle b \rangle \ne 0$
becomes a parameter in the system of 
mean-field equations, and the chemical potential $\mu$ is determined by the
requirement that the excitation spectrum is gapless (Goldstone modes). We find
both long-range order of N\'eel type, where all correlations are
AF only, and spiral order where there are one condensate 
and five bond-order parameters in the solution. The latter number arises 
because all links $J_1$ are equivalent, as are all links $J_2$, {\it i.e.} no 
solutions can be found where there is a symmetry-breaking analogous to the 
dimerization in the MG state, so in general only 2 order parameters are 
required for each type; $J_{2}^{\prime}$ links are AF only, requiring one.
Disordered solutions have no 
Bose condensation, a gapped excitation spectrum and a six-parameter solution
including $\mu$. The phase diagram is shown in Fig. 4. It is gratifying to
find that the AF and spiral ordered phases appear in the expected locations,
and that the phase boundaries are in qualitative agreement with those deduced
from both the dimer and zig-zag chain limits. We note in particular the two
features that i) there is an upper boundary for the regime of spiral order at
sufficiently large $J_2$ and ii) there is a continuous transition directly 
from AF- to spiral-ordered regimes.

	Applying the linear spin wave technique, again (Fig. 4) we find 
regimes with both AF and spiral long-range order, and again there is a 
definite upper limit in $J_2$ on the latter. In this case there is no direct 
transition between the ordered phases, which are always separated by a 
disordered region. This can be shown in detail by studying the ground state 
energy and staggered magnetization. A similar contradiction has been found 
previously \cite{rmpb} in the square lattice with frustrating next-neighbor 
superexchange. For this feature we have no way of distinguishing between 
the phase diagrams of Fig. 4 on the basis of these studies alone, and
leave as open the possibility that the transition between ordered phases may
be continuous along a finite line in parameter space. We may not exclude the
existence of a phase which is none of the four discussed hitherto, as an
intermediate between both the two ordered phases and the two (degenerate and
non-degenerate) gapped phases. 

\begin{figure}[hp]
\centerline{\psfig{figure=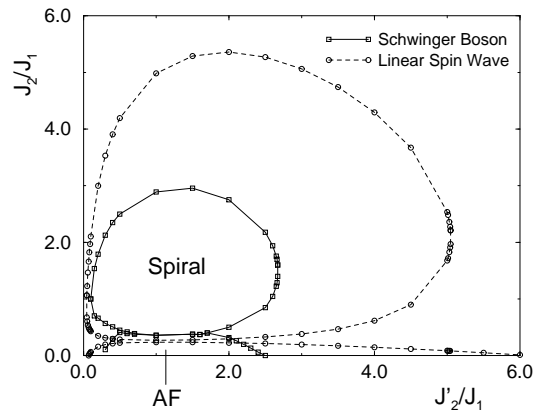,height=6.2cm,angle=270}}
\medskip
\caption{Phase boundaries of the ordered regions, deduced by the Schwinger 
boson and linear spin wave techniques.}
\end{figure}

	All of the approximations applied have tended to overestimate the 
extent of the parent region, although the deduced phase boundaries are in 
good qualitative agreement. The quantitative aspects may be addressed by
numerical techniques. When $J_2$ is zero, the system is an anisotropic 
honeycomb lattice, which has an unfrustrated ordered state for suitably large 
$J_1$ ($J_{2}^{\prime} \ne 0$), and a spin-gap phase otherwise. Because the 
absence of frustration removes the sign problem, the quantum critical point 
on this axis can be found essentially exactly by large-system Quantum Monte 
Carlo studies. The result $J_{2}^{\prime} / J_1 = 1.74$ \cite{rtpc} fixes 
the unknown end of the phase boundary of the non-degenerate spin-gap regime. 
Comparison with the bond-operator (1.15) and Schwinger boson (2.65) results 
indicates the validity of each. We have in addition performed exact 
diagonalization (ED) studies on small systems, from which for the present 
purposes it is possible to locate crossover regions between phases by 
analyzing the quantum numbers of the ground states. This work remains in 
progress, and the results will be presented in detail elsewhere. 

	We conclude with a more schematic discussion of the important limits
of large $J_1$ and large $J_2$. When $J_1$ is large, the system is one of 
weakly coupled chains with a small, marginally relevant next-neighbor 
frustration $J_2$. The action of the interchain coupling $J_2^{\prime}$ has 
been discussed extensively in recent literature for the anisotropic square 
lattice. \cite{raea,rs,rw} Despite the difference in behavior noted above for
this geometry when $J_2 > J_{2c}$, we 
may, following Ref. \onlinecite{rs}, cast the interchain interaction as an 
effective, staggered mean field and deduce that any finite $J_2^{\prime}$
leads to magnetic order while the spectrum of the isolated chains 
remains gapless. Thus for any frustration $J_2 < J_{2c}$, 
the presence of a $J_2^{\prime}$ term can be expected by this argument to
induce AF order. It remains to compute the value of $J_{2c}$ on a frustrated
chain in the presence of the effective staggered field: if this quantity does
not track the phase boundary of the multiply degenerate MG-type ground state, 
we are presented with the existence of an intermediate phase in the region of 
the conformal point and the known phase boundary. This may be the state of no
magnetic order and no spin gap arising in the above discussions, \cite{raea,rw}
which, following the analysis of the ordered states, may also be a candidate 
intermediate regime between AF and spiral order.

	At large $J_2$ the problem is one of chains with competing ladder and 
frustrated couplings. The ladder coupling is relevant, and opens a gap 
$\Delta$ linear in $J_{2}^{\prime} / J_2$ to the nondegenerate state. The 
zig-zag chain coupling opens an exponentially small gap \cite{raw} $\Delta 
\sim \exp (- J_{2} / J_1)$ to the multiply-degenerate state. How each bond 
type affects the gapped state established by the other is unknown, 
as both gaps may close, leaving some intermediate phase, or there 
may be a crossover between the gapped states. In the former case, our results 
provide a strong indication against the possibility that the gapless phase 
is ordered, but not that of a disordered, gapless phase in 2d arising due 
to the competition of the couplings. In the latter case, one may postulate a 
crossover where the singlet-triplet gap changes smoothly, remaining finite
while the low-lying singlet states of the zig-zag chain limit are simply 
split by increasing $J_{2}^{\prime} / J_1$, or may also undergo a 
level-crossing with a higher-lying singlet. Examination of ground-state 
quantum numbers by ED will be particularly useful in resolving this issue.

\begin{figure}[hp]
\centerline{\psfig{figure=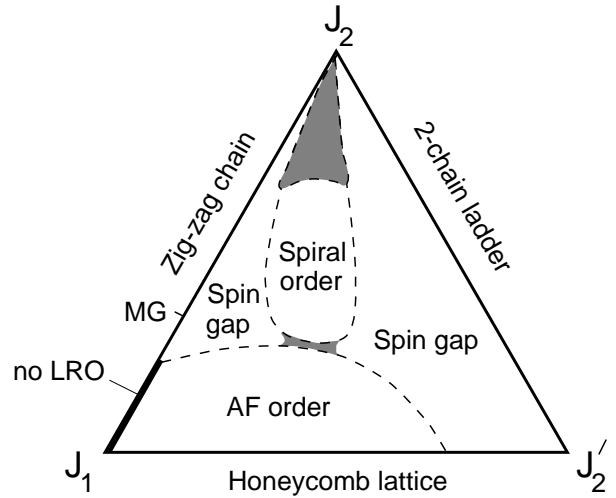,height=6.5cm,angle=0}}
\medskip
\caption{Schematic phase diagram of the frustrated coupled ladder system. }
\end{figure}

	Fig. 5 summarizes the preceding analyses and discussion. The shaded
regions represent those where open questions remain to be addressed. Returning
to the materials which have the frustrated coupled ladder structure, the
original ladder compound SrCu$_2$O$_3$ would appear close to the isotropic
point on the axis $J_1 = 0$, and is well described as a liquid of resonating 
singlets residing primarily on the ladder rungs. That the frustrating
interaction is actually FM is of little consequence here. The zig-zag chain
compound SrCuO$_2$ has very strong $J_2$, and appears close to the upper
vertex in Fig. 5; while $J_2^{\prime}$ is extremely small in this material,
an interchain coupling exponentially small in $J_2/J_1$ is sufficient to 
destroy the MG-type spin-gap state, so the real 
material may be located in the crossover regime of the previous
paragraph. CaV$_2$O$_5$ shows a large spin gap, a result in qualitative 
agreement with expectation for a system with $J_1 \sim J_2 \sim J_2^{\prime}$,
\cite{rmk} as we have seen that the frustrating interaction $J_1$ must be
significantly larger than the ladder terms in order to remove the system from
the non-degenerate gapped phase. 

%\section{Conclusions}

In summary, the frustrated coupled ladder system represents well the wealth of 
interesting physics to be found in low-dimensional quantum magnets, and
provides a valuable means of studying quantum phase transitions. 

%\section{Acknowledgements}

	We are indebted to M. Troyer for the quoted QMC result, and 
acknowledge also useful discussions with J. Bonvoisin, P. Millet and 
D. Poilblanc.

\end{document}